# Comment on "Hydrogen-Antihydrogen Collisions"

For the prototype of matter/antimatter interactions (1a)

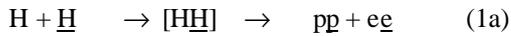
$$H + \underline{H} \rightarrow [H\underline{H}] \rightarrow p\underline{p} + e\underline{e} \quad (1a)$$

Froelich et al. found [1] that $H\underline{H}$, like an atomcule [2], can survive longer than expected. Unfortunately, they did not calculate the PEC (potential energy curve) of intermediate 4-fermion complex $H\underline{H}$, which is very close to that of $H_2$ [3]. This places constraints on matter/antimatter reactions in general and on $H\underline{H}$ in (1a) in particular. A perfect Babel of tongues on conventional "matter" results when considering matter/antimatter reactions like

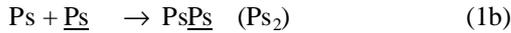
$$Ps + \underline{Ps} \rightarrow Ps\underline{Ps} \quad (Ps_2) \quad (1b)$$

CP-symmetries in neutral N-particle systems (N=2,3,4), with centers of mass $c_M$ and of charges $c_C$, illustrate this:

(i) N=2; short living ($10^{-12}$ sec) charge-conjugated pairs $P_x$ (positronium Ps, protonium Pn); $c_M$ and $c_C$ coincide. Pseudo atom Ps is stable, energy and spectrum derive from Bohr theory. Ps belongs to the "matter" domain, although it consists of equal amounts of matter and antimatter ($e;\underline{e}$). By symmetry, Ps may be classified as antimatter. But if Ps is "matter", *antipositronium* $\underline{Ps}$ is "antimatter" (see below). Mass-asymmetrical pairs ($p;e$ in H, $\underline{p};\underline{e}$ in $\underline{H}$) are stable.

(ii) N=3; charge-conjugated pairs in atomcules; $c_M$ near $c_C$. Antiprotonic helium $He^+\underline{p}$ has a lifetime of μsec, longer than Ps. It is new "matter" [2]. Antiparticle $\underline{He}^-p$ is not (yet) observed. In (i)-(ii), "matter" is a matter/antimatter pair.

(iii) N=4; atom/antiatom pairs as in (1); $c_M$ and $c_C$ do not coincide. Molecules $X\underline{X}$, if stable, are "matter" and building blocks are matter/antimatter pairs. C-symmetry in atom and antiatom is broken by lepton/nucleon mass difference. Annihilation is hampered and lifetimes may equal or exceed that in (ii). Even mass symmetrical N=4 systems like $Ps_2$ in (1b) are believed to be stable [4] (see also below).

If, in reality, "matter" consists of matter/antimatter pairs, this leads to confusion. Neutrality requires 1/1 contributions in terms of charges not of masses. Short lifetimes apply if N=2 and $c_M$ coincides with $c_C$. But simply rotating Ps gives $\underline{Ps}$ due to mass symmetry. Comparing all forms as in (1c)

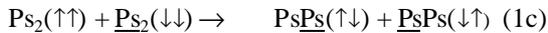
$$Ps_2(\uparrow\uparrow) + \underline{Ps}_2(\downarrow\downarrow) \rightarrow Ps\underline{Ps}(\uparrow\downarrow) + \underline{Ps}Ps(\downarrow\uparrow) \quad (1c)$$

gives symmetric/antisymmetric parallel dipole alignments. Internal C-symmetry in $Ps_2$ becomes a P-symmetry effect, $Ps\underline{Ps}$. By mirror-symmetry and with respect to the $Ps\underline{Ps}$-$c_M$, Ps is left- $Ps_L$ and $\underline{Ps}$ right-handed $\underline{Ps}_R$ (or vice versa). As with Ps, $Ps\underline{Ps}$ is either matter or antimatter. Is it also stable "matter"? If R separates the centers of mass (or charge) and if a is the Ps-radius, interactions V(R) in (1c) for R >> a are

$$V_{\uparrow\downarrow}(R) = -2e^2/R + 2e^2/\sqrt{(R^2 + 2a^2)} = -2e^2a^2/R^3 \quad (2a)$$
$$V_{\uparrow\uparrow}(R) = -V_{\uparrow\downarrow}(R) = (-1)^p V_{\uparrow\downarrow}(R) = +2e^2a^2/R^3 \quad (2b)$$

if parity operator p=1. With (2a), stable "matter" $Ps_2$ [4], not yet observed [5], is matter/antimatter pair $Ps_R\underline{Ps}_L$.

Operator p in 4-fermion Hamiltonians leads generically to singlet-triplet splitting [3]. The drastic effect of Coulomb interactions (2) applies to *4-fermion* complexes in (1), not to the threshold *2-fermion* systems. But, unlike in $Ps_2$, rotation has a different effect on energy than inversion of charges if $c_M$ and $c_C$ do not coincide, as in $H\underline{H}$ [3]. In cold atom region $R/R_e = 20$, difference is 0.01 cm$^{-1}$ in favor of a charge-inverted state [3]. Matter/antimatter pair $H\underline{H}$ in (1a) can also be "matter" with not too short a lifetime [1]. This might explain finally why an $H\underline{H}$-PEC matches the observed $H_2$-PEC [3]. Chemical gaps $\alpha^2\mu c^2$, Coulomb thresholds $e^2/R_e$ for mass asymmetrical charge-conjugated ion pairs $X^+\underline{X}^-$, are $10^{-7}$ times annihilation gaps $m_X c^2$ [3].

The 4 different particles $p, \underline{p}, e$ and $\underline{e}$ in $H\underline{H}$ secure its $c_M$ is *prochiral*, which explains generically why chirality/left-right asymmetry is so important when complex matter is formed from diatomic bonds (bio-molecules, DNA...). With H, this 2$^{nd}$ order stereo-directing effect can only show in *intermolecular* $H\underline{H}$ ("$H_2$") reactions leading to liquid, solid and, eventually, metallic hydrogen.

Nature prefers unconventional charge distributions in N=4 neutral positive mass systems, which result in anti-symmetric matter/antimatter interactions in a generic way and, finally, in stable "matter" [3]. More [3,4] or less [1,2] stable micro-scale antimatter-domains exist close to matter-domains. The universe consists mainly of hydrogen. Stable micro-scale "matter" $H\underline{H}$ gives matter/antimatter symmetry for the universe on the macro-scale [3]. This avoids having to look for large-scale antimatter-domains at large distances (Mpc) from matter-domains [6] to conserve this symmetry.

G. Van Hooydonk,
   Departments of Library Sciences and of Physical & Inorganic Chemistry, Ghent University, Rozier 9, Ghent, Belgium